# The effect of the noncentral impurity-matrix interaction upon the thermal expansion and polyamorphism of solid CO – $C_{60}$ solutions at low temperatures.

A.V. Dolbin[1], V.B. Esel'son[1], V.G. Gavrilko[1], V.G. Manzhelii[1], N.A. Vinnikov[1],
G.E. Gadd[2], S. Moricca[2], D. Cassidy[2], B. Sundqvist[3].

[1] Institute for Low Temperature Physics & Engineering NASU, Kharkov 61103, Ukraine
[2] Australian Nuclear Science & Technology Organisation, NSW 2234, Australia
[3] Department of Physics, Umea University, SE - 901 87 Umea, Sweden

Electronic address: dolbin@ilt.kharkov.ua

**Abstract.**

Orientational glasses with CO molecules occupying 26% and 90% of the octahedral interstitial sites in the $C_{60}$ lattice have been investigated by the dilatometric method in a temperature interval of 2.5 – 23 K. At temperatures 4 – 6 K the glasses undergo a first-order phase transition which is evident from the hysteresis of the thermal expansion and the maxima in the temperature dependences of the linear thermal expansion coefficients $\alpha(T)$, and the thermalization times $\tau_1(T)$ of the samples. The effect of the noncentral CO – $C_{60}$ interaction upon the thermal expansion and the phase transition in these glasses was clarified by comparing the behavior of the properties of the CO – $C_{60}$ and $N_2$ – $C_{60}$ solutions.

**Introduction**

Below T=90 K fullerite $C_{60}$ transforms into an orientational glass. According to dilatometric and X-ray structural data [1 – 7], the gases dissolved in $C_{60}$ produce a significant effect on the thermal expansion of the glass and cause a first-order phase transition (polyamorphism) in it. It is interesting to find out how particular molecular parameters of the admixture gas can influence the properties of $C_{60}$ lattice as a result of a impurity-matrix interaction. To judge accurately the effect of varying a certain molecular parameter, the gas impurities should be chosen so that they differ mainly in this particular parameter, whilst other molecular parameters that may have an effect on the $C_{60}$ lattice are essentially kept the same. For example, we would like to probe the effect on the impurity-matrix interaction of altering the electronic charge distribution within a diatomic gas. The choice of a homo and hetero diatomic gas with similar molecular bond lengths would be a good starting point for investigating this important question. We have conveniently chosen CO – $C_{60}$ and $N_2$ – $C_{60}$ solutions. In contrast to $O_2$, CO and $N_2$ molecules do not react chemically with $C_{60}$ at the temperatures to which $C_{60}$ has to be heated to desorb volatile impurities. These molecules also have practically identical molecular weights (M(CO)=28.0105, M($N_2$)=28.0134) as well as comparable gas-kinetic diameters ($\sigma$(CO)=3.766Å, $\sigma(N_2)$=3.756Å) [8], but they differ significantly in electric quadrupole moments Q (Q(CO)= -2.839 · $10^{-26}$ esu, Q($N_2$)= - 1.394 · $10^{-26}$esu) [9]. $N_2$ also does not have a dipole moment whereas CO does. However, as will be discussed further on, it is the quadrupole moment and not the dipole moment, that contributes most to the effect that these impurities have upon the low-temperature thermal expansion and polyamorphism of $C_{60}$.

The dilatometric data on orientational $C_{60}$ glasses with molar $N_2$ concentrations ($N_2$-to-$C_{60}$ molecule ratio) of 9.9% and 100% has previously been published in [3], so that this paper extends the studies to include those from CO – $C_{60}$ solutions, followed by comparison of the two data sets.

In this study, we investigate the impurity effect of CO on the properties and phase transformations of orientational $C_{60}$ glasses. Solutions of CO – $C_{60}$ with both 26 mol. % CO and 90 mol. % CO, were investigated.

The impurity ($N_2$, CO) molecules occupy the octahedral interstitial cavities in the $C_{60}$ lattice, of which there is effectively one octahedral cavity per $C_{60}$ molecule. As a result of this, the molar CO and $N_2$ concentrations are equal to the $N_2$ and CO occupancies of the octahedral sites in the $C_{60}$ lattice.

**Samples and measuring technique**

The $C_{60}$ sample with 26 mol. % CO was prepared as follows. Prior to saturation with CO, the sample, which was a pressed cylinder of solid $C_{60}$ powder, 9 mm high and 10 mm in diameter (prepared by a procedure as described in reference [2]), was kept for 72 hours under dynamic evacuation to remove gas impurities (P=1·$10^{-3}$ mm Hg, T=400°C). The outgassed sample and cell, was filled with CO gas at room temperature to a pressure of 760 mm Hg and sealed. The sample was kept under these sealed conditions for 105 days.

The thermal expansion of the CO – $C_{60}$ solutions was investigated using a low temperature capacitance dilatometer. Its design and the measurement technique are detailed in [14].

Immediately before the dilatometric measurement, the measuring cell with the CO – $C_{60}$ sample and which was filled with CO, was cooled slowly to 65K, which is just below the freezing point of CO at 68 K. The cell was evacuated at this temperature to remove the condensed CO, that was CO that had not been absorbed by the sample. The sample was pumped on further until a base pressure of 1 x· $10^{-5}$ mm Hg was attained, followed by cooling of the sample to the base temperature of 4.2K. The thermal expansion of the CO – $C_{60}$ sample was measured after a four-hour exposure to this temperature.

After measuring the thermal expansion of the sample, the amount of gas impurities and their compositions were determined qualitatively and quantitatively using a vacuum desorption gas analyzer [12]. It was found that about 26% of the octahedral cavities of the $C_{60}$ lattice were occupied by CO. Most of the CO was desorbed on heating the sample to 300°C (Fig. 1). The preparation and analysis techniques for the $C_{60}$ sample with 90 mol. % CO are described in [13].

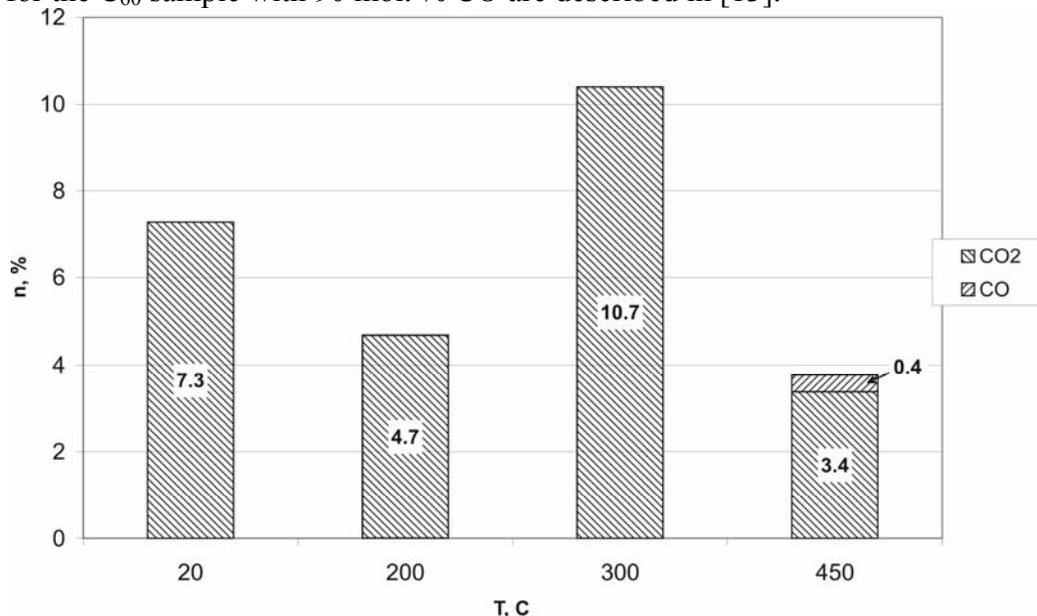

Fig. 1. The composition of the gas mixture (in percentage of occupation of the octahedral cavities) desorbed from the $C_{60}$ sample with 26 mol. % CO on stepwise heating the sample to T=450°C.

**Results and discussion**

The temperature dependences of the linear thermal expansion coefficient (LTEC) $\alpha(T)$ of pure $C_{60}$ and of the $C_{60}$ samples with different contents of the CO impurity are shown in Fig. 2. The $\alpha(T)$-values are averaged over several measurement series. Owing to the cubic symmetry of their lattices, the thermal expansion of the samples can be described with a single LTEC.

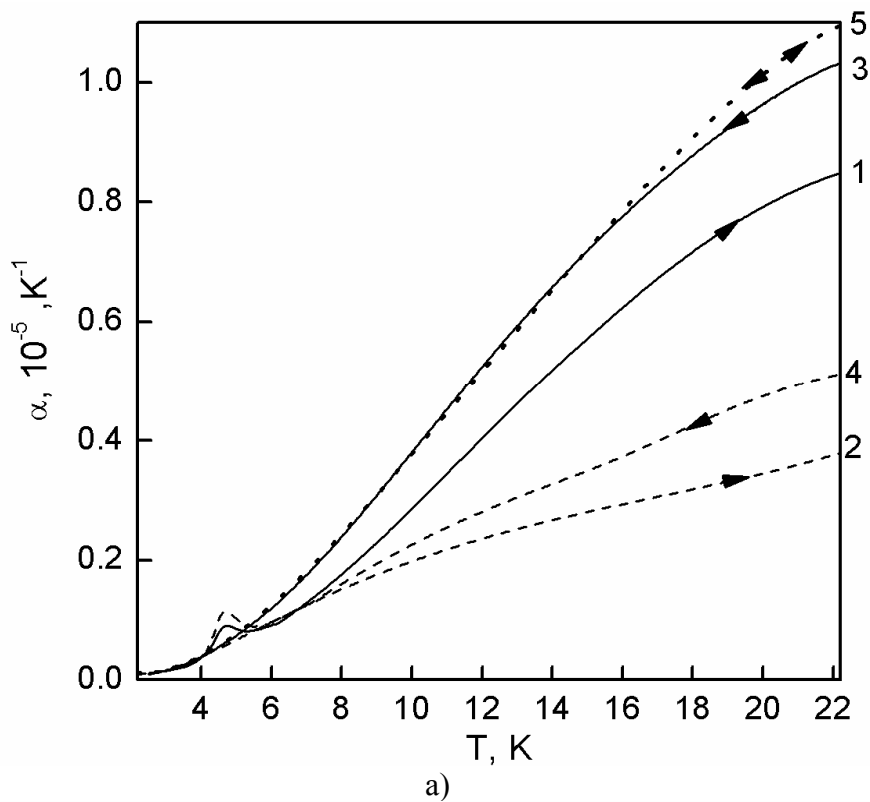

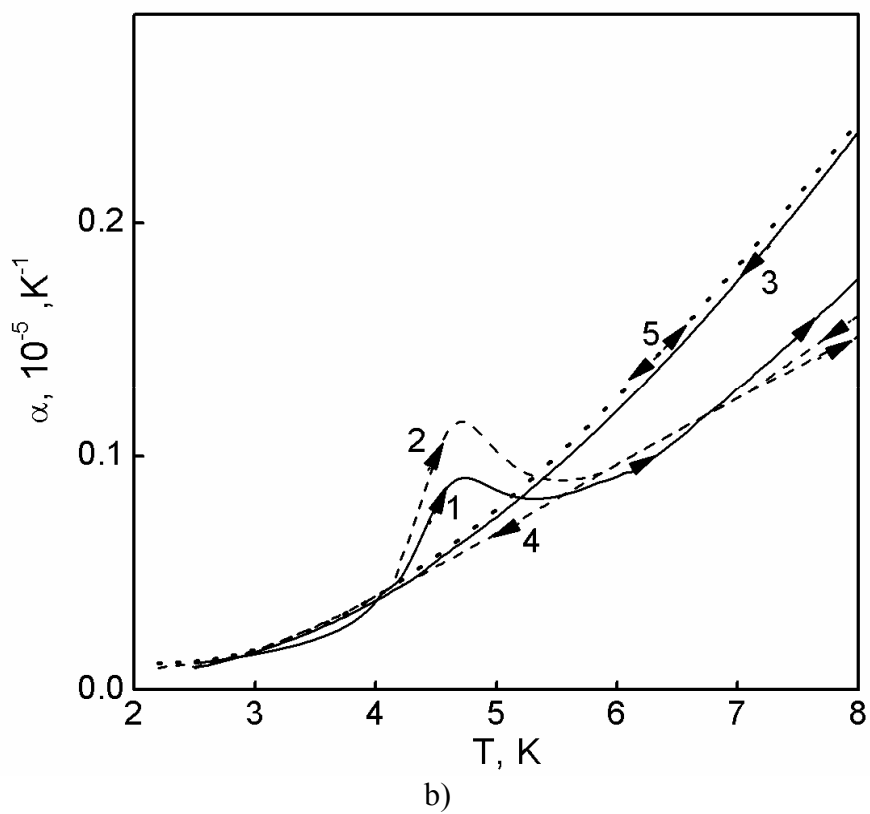

Fig. 2. The temperature dependence of the linear thermal expansion coefficient of CO – $C_{60}$ solutions in temperature intervals (a) 2.5 – 23 K, and (b) 2.5 – 8 K. The curves numbered 1 and 2 are from heating the 26 mol. % and 90 mol. % CO-$C_{60}$ samples respectively, whilst curves 3 and 4 are from cooling the 26 mol. % and 90 mol. % CO-$C_{60}$ samples respectively. The dotted line (5) is from pure $C_{60}$ by either heating or cooling the sample.

The thermal expansion of the investigated samples exhibited a number of specific features. On heating (curves 1, 2) and subsequent cooling (curves 3, 4) the thermal expansion coefficient has a hysteresis which points to a first-order phase transition in the orientational CO – $C_{60}$ glasses. No hysteresis was observed for pure $C_{60}$ (curve 5). From the two different CO –$C_{60}$ samples, it appears that the onset of the hysteresis shifts towards higher temperature with increased impurity concentration, shifting from 3 K for the 26 mol. % CO to 4 K for the 90 mol. % CO. Within the temperature range starting from the lowest measured temperature of 2.5 K to the respective hysteresis onset temperatures, it is found that the α(T) for a particular sample are practically identical for both the heating and cooling curves. Moreover, the LTECs of both the CO – $C_{60}$ samples (26 and 90 mol. % CO) and that of pure $C_{60}$ also coincide within the measurement error. On heating in the interval 4 – 6 K there is a region of instability with higher experimental errors and local LTEC maxima (Fig. 2), however, the errors are appreciably lower than the maxima heights observed. It is assumed [3 – 5] that within this interval of temperatures, occurs the first-order phase transition between the two differing impurity doped orientational glasses.

Previous investigations show [1 – 5] that the thermal expansion of gas-doped $C_{60}$ contains positive and negative components with different characteristic relaxation times ($\tau_1$ and $\tau_2$, respectively). With a temperature change of the sample, the positive component is attributed [1 – 5] to the process of temperature equalization over the bulk sample (thermalization), whilst the negative component accounts orientational changes of the $C_{60}$ molecules induced by the temperature change of the sample. Since a $C_{60}$ crystal is perceived as consisting of domains of different orientational orders of the $C_{60}$ molecules, and with these domains in a particular crystal separated by interlayers of $C_{60}$, it has been concluded theoretically [15-17] that the negative component of the thermal expansion observed at these low temperatures studied, results from the $C_{60}$ reorientation within these actual interlayers and not in the domains themselves.

The thermal expansion of $C_{60}$ samples doped with CO also has two components. They were separated in a similar fashion to the techniques described in [1]. The temperature dependences of the positive and negative components for samples with different CO concentration are illustrated in Fig. 3.

It is seen from Fig. 2(b) that all the LTEC curves are lower than the LTEC of pure $C_{60}$ over this temperature range, with the heating and cooling curves of the 90 mol. % CO and the heating curve of the 26 mol. % CO, being markedly lowered. From Fig. 2(b) we can conclude that this lowering scales with increases in the concentration of the CO impurity. Above 8K, although the heating and cooling LTEC curves for the 90 mol. % CO and the heating curve of the 26 mol. % CO appear lowered even further than the corresponding LTEC curve of pure $C_{60}$, the cooling curve of the 26 mol. % CO is more or less identical with that of pure $C_{60}$. However if we consider just the positive component to this curve, as shown in Fig. 3, it is seen to also be lower than that of pure $C_{60}$ (which only exhibits a positive component). In $N_2$-$C_{60}$ solutions this lowering effect of the LTEC curves, exists only at high $N_2$ concentrations and is much less [3]. This lowering effect is explained as follows. Over the temperature range spanned in our experiments (2.5 - 23 K), the thermal expansion of pure $C_{60}$ is determined by the changes with temperature that occur in a range of phenomena, which predominantly include the translational lattice vibrations, the $C_{60}$ librations and the soft modes and two-level systems of the $C_{60}$ glasses, and in particular those associated with the changing of the relative orientations of the $C_{60}$ molecules with respect to each other. The admixed gas molecules within the octahedral sites can affect the above contributors as well as making their own contribution to the thermal expansion of the solid CO-$C_{60}$ solution, through its own thermal motions.

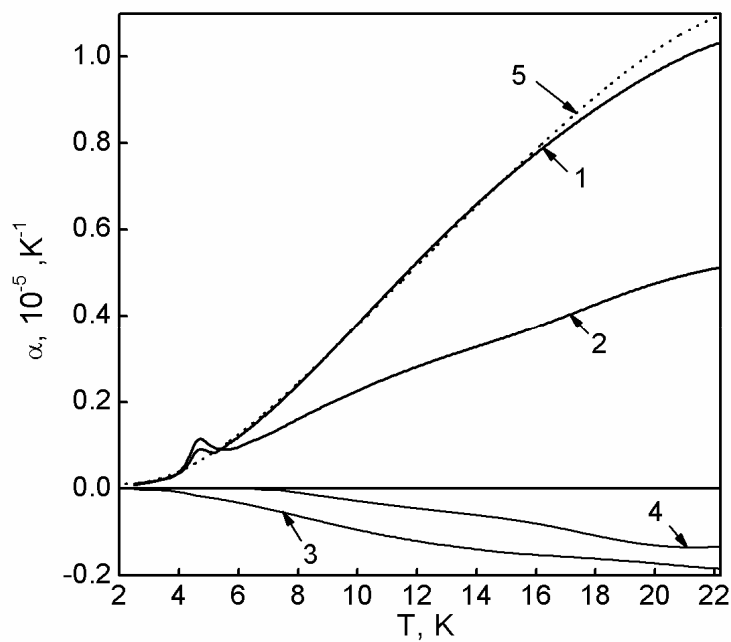

a)

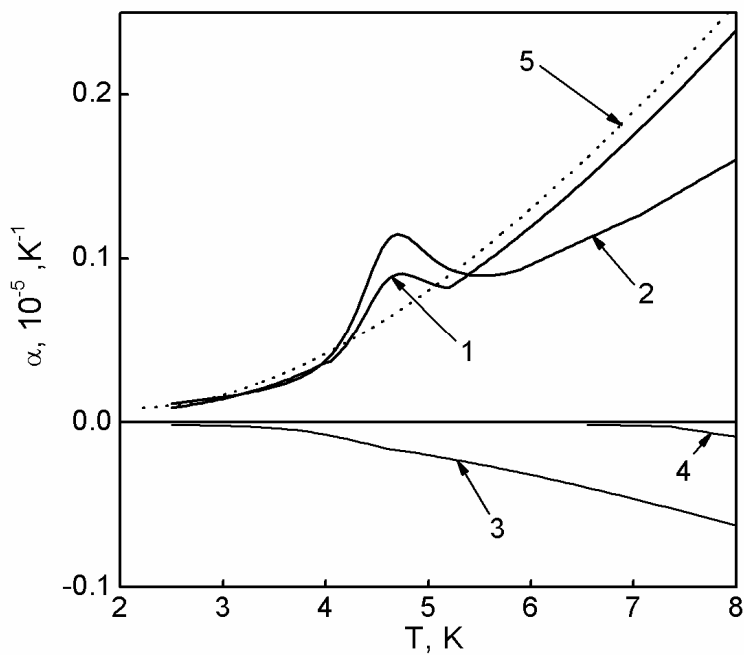

b)

Fig. 3. The temperature dependences of the positive and negative components of the thermal expansion coefficient of CO-$C_{60}$ solutions as studied in the intervals of (a) 2.5 – 23 K, and (b) 2.5 – 8 K. The positive contributions are the labelled curves 1 – 26 mol. % CO, and 2 – 90 mol. % CO, whilst the negative contributions are labelled as the curves: 3 – 26 mol. % CO, and 4 – 90 mol. % CO. Pure $C_{60}$ which only exhibits a positive contribution is shown as a dotted line (5).

As noted above, the thermal expansion coefficients of pure $C_{60}$ and the CO-$C_{60}$ solutions coincide at the lowest temperatures of the experiment. This means that at these temperatures the CO impurity has little effect on the dominant contributors to the thermal expansion that being the translational lattice vibrations, the two – level systems and the soft modes [18]. The weak effect of the impurity on the translational vibrations of the $C_{60}$ lattice is quite natural because CO adds little to the effective molecular weight of the CO-$C_{60}$ solutions and changes the lattice constant of $C_{60}$ at most by 0.15 % [19]. As the temperature rises, the contributions of the $C_{60}$ librations and the motions associated with the CO molecule (translational, rotational and internal vibrational) increase significantly. But any contribution from the CO through its increased thermal motion can only lead to higher LTEC – values. Therefore, the lower LTECs of the CO-$C_{60}$ solutions in comparison with those of pure $C_{60}$ must be attributed to the diminished contribution of the $C_{60}$ librations. This arises from the fact that the CO molecules at $T \leq 77$ K are oriented in a particular fashion within the octahedral interstitial sites of $C_{60}$ [11, 20, 21] so that there is a noncentral interaction between the impurity and the surrounding $C_{60}$ molecules and which is not nullified by any rotation of the CO molecules within the sites. The CO molecule has both dipole and quadruple electrical moments although the dipole moment is rather weak [22], so that the noncentral CO-$C_{60}$ interaction is mainly determined by the quadrupole moment of the CO molecule. This CO orientational induced noncentral force interaction acting on the $C_{60}$ molecules tends to increases the frequency of their librations. As a result, the contribution of the $C_{60}$ librations to the positive component of the thermal expansion for these CO-$C_{60}$ solutions, and over the studied temperature range, is reduced as compared to that of the pure $C_{60}$ sample. Only at higher temperatures will their effect be realised as their contribution increases with temperature as they become more and more thermally activated. This effect is weaker for the $N_2$-$C_{60}$ solutions as the quadrupole moment of $N_2$ molecules is much smaller and this is clearly seen in Fig. 4 that compares the positive and negative components of the thermal expansion for samples with 90 mol. % CO and 100 mol. % $N_2$. We have chosen these two samples to compare, because the effects of impurities upon the thermal expansion of doped fullerites are most evident at their high concentrations.

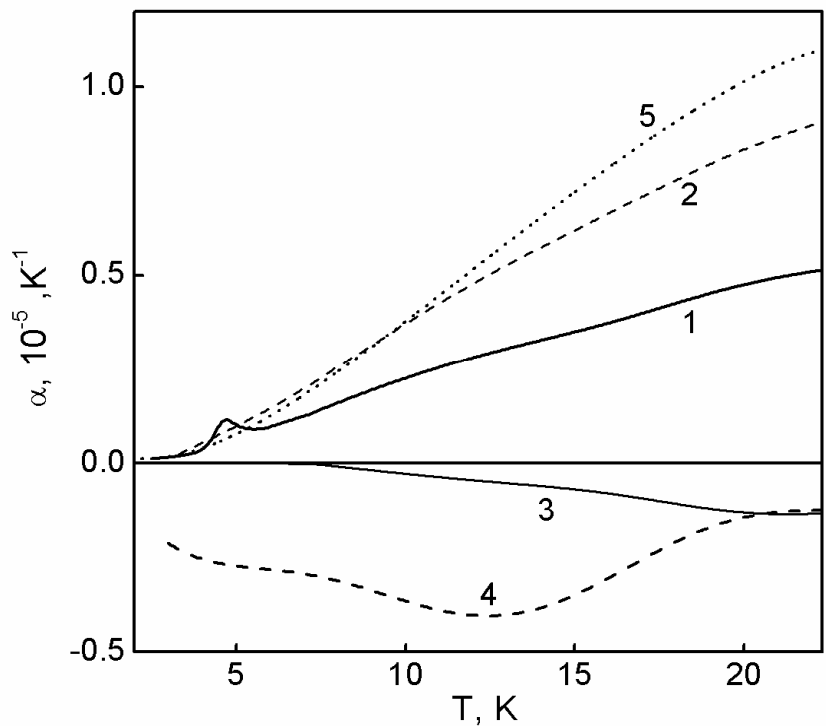

Fig. 4. The temperature dependences of the positive and negative components of the linear thermal expansion coefficient are shown for both CO-$C_{60}$ (solid lines) and $N_2$-$C_{60}$ (dashed lines) [3] solutions. The positive components are the curves: 1 – 90 mol. % CO, and 2 – 100 mol. % $N_2$, whilst the negative components are the curves 3 – 90 mol. % CO, and 4 – 100 mol. % $N_2$. Again pure $C_{60}$ which only has a positive contribution is shown as the dotted line (5).

In the context of the above consideration, the negative component of the thermal expansion is determined by the probability of reorientation of the $C_{60}$ molecules in the domain interlayers. It is found that the magnitude of the negative component of the LTEC for the CO-$C_{60}$ solutions decreases considerably as the CO concentration increases from 26% to 90% (Fig. 3). The absence of a negative component in the thermal expansion of pure $C_{60}$ prompts us to conclude that on dissolution of CO in the $C_{60}$ lattice, the probability of $C_{60}$ reorientation in the domain interlayers must first increase with impurity concentration up to a certain impurity concentration but increasing the impurity concentration past this, it must start to decrease again and Fig. 3 even suggests it contribution is reduced to zero contribution at 100% occupancy with CO. By contrast, for $N_2$-$C_{60}$ solutions, in which the noncentral interaction between the $N_2$ and the $C_{60}$ molecules is weaker, there is an opposite trend with the negative contribution to the thermal expansion being much greater at higher $N_2$ concentrations than at lower ones. It should be noted that a change from a low to a high impurity concentration will reduce the temperature interval of the negative contribution for the CO-$C_{60}$ solution but in contrast increases it for the $N_2$-$C_{60}$ solution [3].

It is of our opinion that these observations of the thermal expansion behaviour for the CO-$C_{60}$ and $N_2$-$C_{60}$ solutions suggests that in both cases, there is a competition between two contrasting mechanisms. On the one hand, we have the CO and $N_2$ impurities introduced into the interstitial sites of $C_{60}$ pushing the neighboring $C_{60}$ molecules farther apart. This mechanism tends to reduce the effect from the non-central interaction between the $C_{60}$ molecules but promote their reorientation. This increases the negative contribution to the thermal expansion. On the other hand, their introduction also results in a noncentral interaction between the impurity and the neighboring $C_{60}$ molecules that reduces the probability of $C_{60}$ reorientation and decreases the negative contribution to the thermal expansion. The first mechanism dominates in the $N_2$-$C_{60}$ solutions while the other prevails in the CO-$C_{60}$ solutions with high CO concentrations.

It is expected that the noncentral interaction between the impurity and $C_{60}$ matrix can affect the characteristic time of $C_{60}$ reorientation ($\tau_2$) within the interlayers between the domains. As seen in Fig. 5, where the $\tau_2$ –values have been extracted from the negative component of the LTEC, are much longer for the CO-$C_{60}$ solution than for the $N_2$-$C_{60}$ one, indicating that the CO molecules with the substantially larger quadrupole moment than that of $N_2$ greatly depress on account of this enhanced non central CO-$C_{60}$ interaction, the probability of $C_{60}$ reorientation.

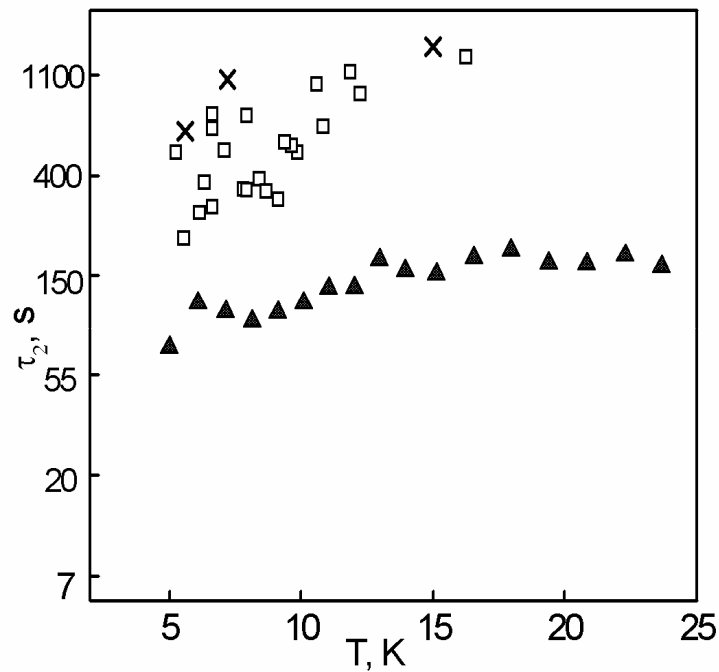

Fig. 5. The characteristic time ($\tau_2$) for $C_{60}$ reorientation extracted from the negative components of the thermal expansion: CO-$C_{60}$ with 90 mol. % CO (×), and 26 mol. % CO (□); and $N_2$-$C_{60}$ with 100 mol. % $N_2$ (▲).

The local maxima in the temperature dependences of the positive components of the LTECs for the CO-$C_{60}$ samples may indicate that within the interval of 4 ÷ 5,5K occurs the temperatures associated with the phase transformations between the orientational CO-$C_{60}$ glasses. This assumption is supported by the analysis of the temperature dependences of the relaxation time $\tau_1(T)$ associated with thermal equilibration of the CO-$C_{60}$ solution, and obtained from the positive component of the LTEC. As shown in Figure 6, this extracted thermalization time $\tau_1$ of the sample, increases sharply in the temperature interval of the local LTEC maxima because the heat supplied to the sample during heating is partially consumed by the phase transformation in the orientational glass.

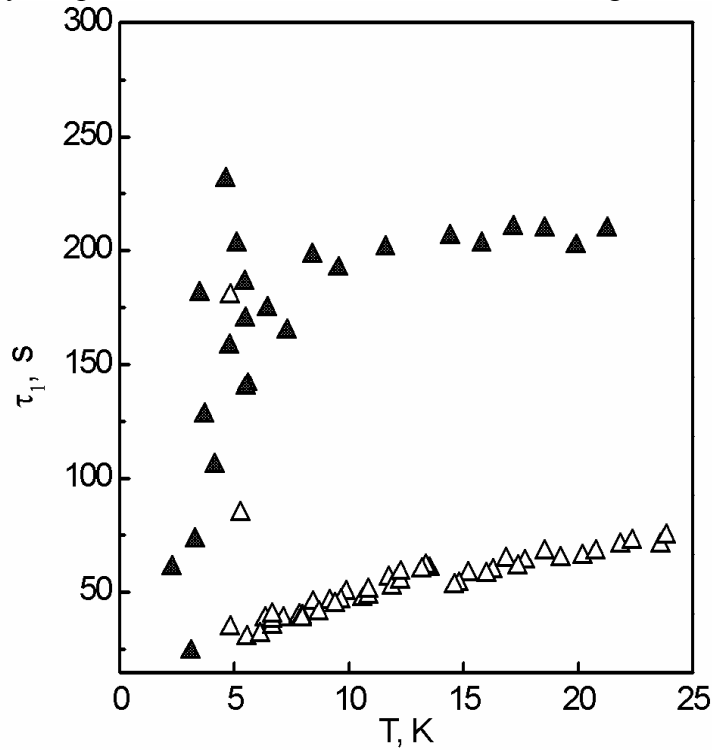

Fig. 6. The characteristic times $\tau_1$ of the positive components of thermal expansion from:
▲ – 90 mol. % CO-$C_{60}$, and △ – 26 mol. % CO-$C_{60}$.

In contrast to the CO-$C_{60}$ solutions, the dependences $\alpha(T)$ and $\tau_1(T)$ of the $N_2$-$C_{60}$ solution have no distinct maxima. The glasses coexisting in gas – fullerites solutions differ only in the orientational order of the $C_{60}$ molecules [2, 15, 16, 17]. Since the noncentral interaction between the impurity and matrix molecules is stronger in the CO-$C_{60}$ solution, we can assume that the latent heat of the phase transformation between the glasses and associated with this change in the orientational order is much larger for this solution, up as maxima in the $\alpha(T)$ and $\tau_1(T)$ plots.

It is known that gas impurities of high concentrations can often cause microcracking and even fracture of the $C_{60}$ samples [2, 3, 5, 23]. The higher $\tau_1$-values for the sample with the high CO concentration can be attributed to evidence for microcracks occurring within such samples. Such an occurrence will increase the thermal resistance and hence the characteristic time of thermalization $\tau_1$.

The phase transformation in the CO-$C_{60}$ samples was investigated by conducting a series of experiments involving thermocycling of the samples at T>5,5K. The thermocycling was performed in several narrow intervals, with the step being no more than 2K. The experimental technique and data processing have been detailed elsewhere [2, 3]. In the course of the thermocycling, the hysteresis loop was found to narrow gradually (from cycle to cycle) and in doing so the negative component of the thermal expansion decreased until the LTECs measured on heating were approaching those obtained on cooling. The process of thermocycling thus brought the system to a more advantageous thermodynamic state between the co-existing glasses. The characteristic times $\tau'$ of this process are shown in Fig. 7 as a function of the average thermocycling temperature.

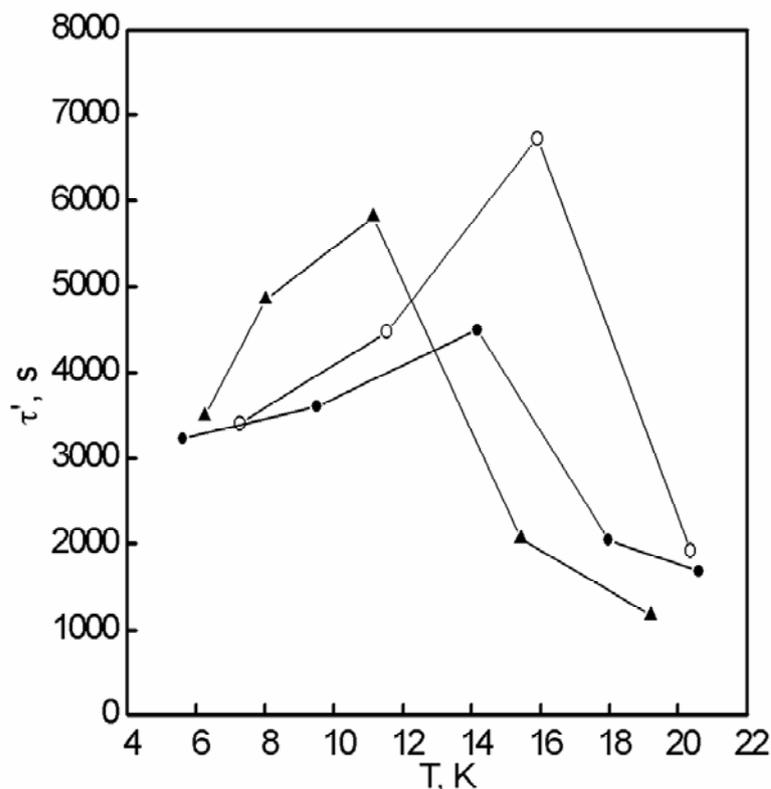

Fig. 7. The temperature dependences of the characteristic time ($\tau'$) for the phase transformation between the orientational glasses for $N_2$- $C_{60}$ with 100 mol. % $N_2$ (▲, [3]) and for CO-$C_{60}$ with, 26 mol. % CO (o), and 90 mol. % CO (•)

It is interesting that the dependences $\tau'(T)$ are similar qualitatively for all three samples, although the $\tau'(T)$ maxima for the CO-$C_{60}$ solutions are shifted towards a higher temperatures. Currently, the lack of information concerning the actual distinctions between the orientational glasses coexisting in the CO-$C_{60}$ and $N_2$-$C_{60}$ solutions, impedes any further analysis of the temperature dependences of $\tau'(T)$ for these systems.

**Conclusions**.

A first order phase transition was observed in the orientational $C_{60}$ glasses at liquid helium temperatures, during dilatometric investigations on two CO-$C_{60}$ solutions with 26 mol. % CO and 90 mol. % CO. The phase transformation revealed itself through observation of a hysteresis in the thermal expansion, the occurrence of local maxima in the temperature dependence of the linear thermal expansion coefficients, and lastly by a maximum in the temperature dependence of the thermalization time $\tau_1$ of the investigated systems. From the temperature range of the observed maxima in the thermal expansion, the phase transitions in the orientational glasses of the CO-$C_{60}$ solutions is believed to occur in the interval 4÷6K.

The thermal expansion of the CO-$C_{60}$ solutions is a sum of positive and negative components and each with the characteristic relaxation times $\tau_1$ and $\tau_2$, respectively. $\tau_1$ is the time of temperature equalization over the sample (thermalization) whilst $\tau_2$ specifies the time of $C_{60}$ reorientation in the interdomain space within the CO-$C_{60}$ crystallites.

We compared the thermal expansion of CO-$C_{60}$ and $N_2$-$C_{60}$ solutions in which the impurity molecules have close gas – kinetic diameters and molecular weights, but where CO has a considerably larger quadrupole moment than $N_2$.

Because of the stronger noncentral interaction between the interstitial CO and the neighboring $C_{60}$ molecules, the CO-$C_{60}$ solution has some specific features that distinguish it from the $N_2$-$C_{60}$ solution. These are:

(i). The linear thermal expansion coefficients (LTECs) are lower in the "high-temperature" phase in comparison with the LTECs of pure $C_{60}$. This is because the frequencies of $C_{60}$ librations are increased through this noncentral interaction and their contribution to the LTECs shifts to temperatures above the T-interval of the experiment.

(ii). The dependences $\alpha(T)$ and $\tau_1(T)$ have maxima in the temperature interval of phase transformation. No maxima were detected in the $N_2$-$C_{60}$ solutions.

(iii). The $C_{60}$ molecules have much longer reorientation times $\tau_2$, which is an obvious consequence of the enhancement of the noncentral interaction between the impurity and matrix molecules.

(iv). There is a change in the concentration dependence of the negative contribution to the LTEC. Two contrasting mechanisms are responsible for these observations associated with the negative LTEC contribution. On the one hand, impurities increase the spacings between the $C_{60}$ molecules which depresses their noncentral interaction and increases the probability of their reorientation. On the other hand, the noncentral interaction between the impurity and matrix molecules decreases the probability of $C_{60}$ reorientation. The first mechanism is dominant in the $N_2$-$C_{60}$ solutions, whilst the other predominates in the CO-$C_{60}$ solutions with higher CO concentrations.

The authors thank Prof. A.S. Bakai for helpful discussions. The authors are also indebted to the Science and Technology Center of Ukraine (STCU) for the financial support of this study (Project Uz-116).